\documentclass[journal]{IEEEtran}

\usepackage{algorithm}
\usepackage{algorithmic}
\usepackage{amsfonts}
\usepackage{multirow}
\usepackage{amssymb}
\usepackage{amsmath, amsthm}
\usepackage{mathtools}
\usepackage{bbm}
\usepackage{graphicx}
\usepackage{epstopdf}

\usepackage{epsf}
\usepackage{amsmath, amsfonts}
\usepackage{latexsym}
\usepackage{caption}
\usepackage{subcaption}
\usepackage{multirow}
\usepackage{multicol}
\usepackage{amssymb}
\usepackage{cite}
\usepackage{float}
\usepackage{enumerate}
\usepackage{color}
\usepackage{scrextend}
\usepackage{refcount}
\usepackage{bbm}
\usepackage{lipsum}
\usepackage{mathtools}
\usepackage{cuted}

\allowdisplaybreaks

\begin{document}

\title{\fontsize{20}{24}\selectfont Consensus-Before-Talk: Distributed~Dynamic~Spectrum~Access via Distributed Spectrum Ledger Technology}

\author{
Hyowoon~Seo, $^\dagger$Jihong~Park, $^\dagger$Mehdi~Bennis, and~Wan~Choi
\thanks{H. Seo and  W. Choi are with the School of Electrical Engineering, Korea Advanced Institute of Science and Technology (KAIST), Daejeon 34141, Korea (e-mail: hyowoonseo@kaist.ac.kr, wchoi@kaist.edu).}
\thanks{$^\dagger$J. Park and $^\dagger$M. Bennis are with the Centre for Wireless Communications, University of Oulu, Oulu 90014, Finland (e-mail: jihong.park@oulu.fi, mehdi.bennis@oulu.fi).}
}




\maketitle

\begin{abstract}
This paper proposes Consensus-Before-Talk (CBT), a spectrum etiquette architecture leveraged by distributed ledger technology (DLT). In CBT, secondary users' spectrum access requests reach a consensus in a distributed way, thereby enabling collision-free distributed dynamic spectrum access. To achieve this consensus, the secondary users need to pay for the extra request exchanging delays. Incorporating the consensus delay, the end-to-end latency under CBT is investigated. Both the latency analysis and numerical evaluation validate that the proposed CBT achieves the lower end-to-end latency particularly under severe secondary user traffic, compared to the Listen-Before-Talk (LBT) benchmark scheme.
\end{abstract}


%
\IEEEpeerreviewmaketitle

\newtheorem{mylemma}{Lemma}
\newtheorem{myremark}{Remark}
\newtheorem{mytheorem}{Theorem}
\newtheorem{mydef}{Definition}
\newtheorem{mycor}{Corollary}
\newtheorem{myexample}{Example}
\newtheorem{mydefinition}{Definition}
\newtheorem{myproposition}{Proposition}
\newcounter{mytempeqncnt}

\newcommand{\tred}[1]{{\textcolor{red}{#1}}}
\newcommand{\tblue}[1]{{\textcolor{blue}{#1}}}

\section{Introduction}

Unlicensed spectrum bands are envisaged to be at the cusp of the collapse, due to the unprecedented overuse by a huge number of WiFi devices \cite{Cisco2016} as well as by the contribution from cellular devices such as licensed assisted access (LAA) \cite{Rupasinghe2014, Hong2011}. Their number of access requests may become too large to be supported by traditional unlicensed spectrum etiquettes that include Carrier Sensing Multiple Access with Collision Avoidance (CSMA/CA) \cite{Bianchi1996, Ziouva2002} and Listen-Before-Talk (LBT) \cite{Song2016, Ko2016, Leu2006} for standalone WiFi and WiFi-cellular coexistence scenarios, respectively. 

Furthermore, even with the unlicensed spectrum, a cellular-grade latency guarantee is expected to be demanded, in order for WiFi connections to provide seamless WiFi-to-cellular experiences. A compelling example could be an ultra-reliable and low-latency communication (URLLC) scenario where a target latency constraint should be ensured anytime anywhere \cite{PetarURLLC:17,MehdiURLLC:18,UR2Cspaswin:17}, regardless of the connection types. This void is difficult to be filled by the traditional spectrum etiquettes that commonly incur random back-off delays due to collisions~\cite{Bianchi1996, Ziouva2002}, as illustrated in Fig.~1-a.

In order to resolve the aforementioned spectrum etiquette problems, by leveraging distributed ledger technology (DLT) we propose a novel unlicensed spectrum etiquette, \emph{Consensus-Before-Talk (CBT)}. In~CBT, the unlicensed users' spectrum requests are first come to a consensus in a distributed way, yielding a \emph{distributed spectrum ledger (DSL)}. The DSL, stored at each user, contains a consensual sequence of the spectrum requests. This sequence is ordered by a pre-defined consensus policy, thereby enabling the distributed dynamic spectrum access of the users without any collision, as shown in Fig.~1-b. 

To enjoy this benefit, CBT needs to pay for the extra consensus latency. For the purpose of minimizing this latency without suffering from severe interference, inspired by the Hashgraph algorithm \cite{bi:Hashgraph}, the users in CBT exchange their spectrum access requests using a gossip protocol, and achieve their consensus by the local computation at each user. In this paper, the resulting end-to-end CBT latency is investigated, and its effectiveness is highlighted by comparing it with a benchmark LBT scheme via analysis and numerical evaluations, followed by the discussion for possible extensions to the proposed CBT. 
 
\begin{figure}
\includegraphics[width=\columnwidth]{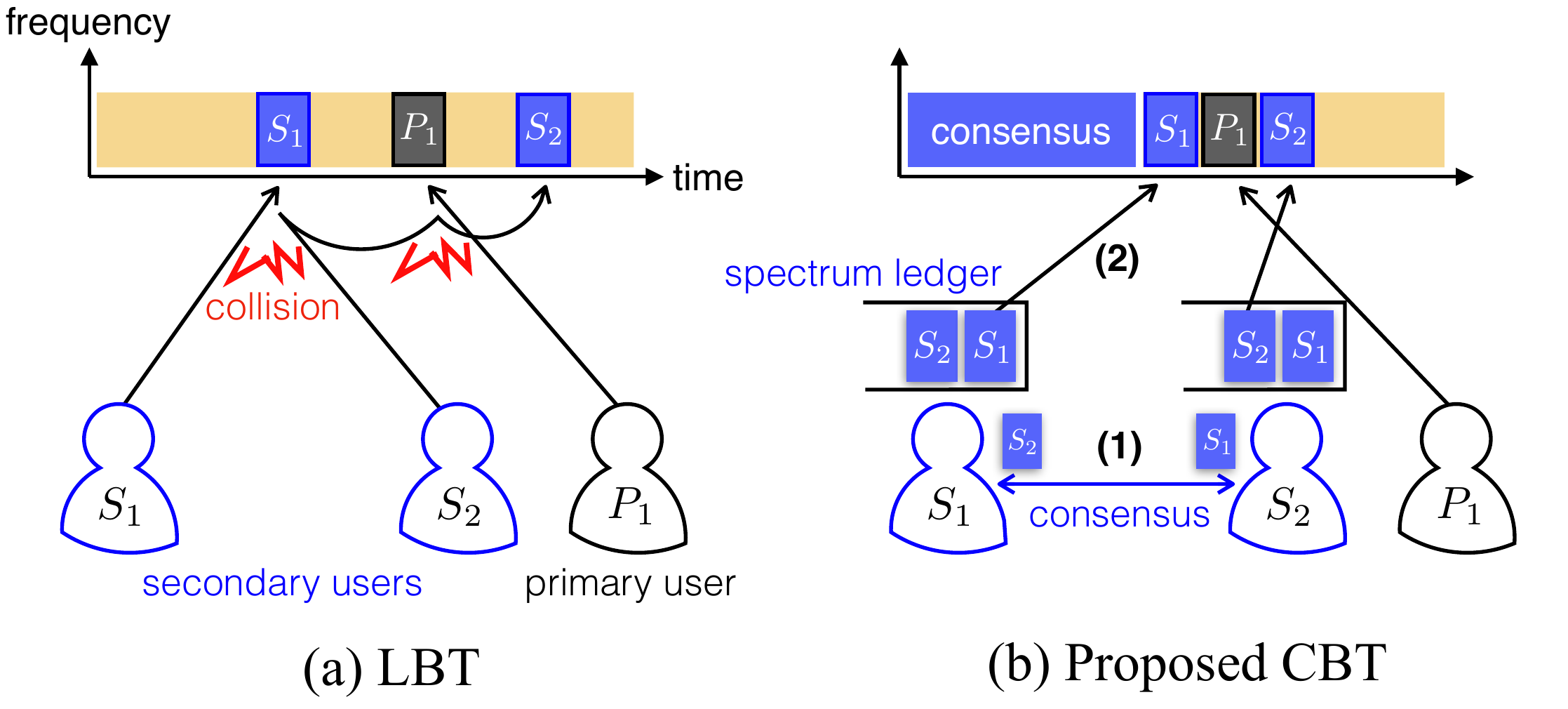}
\caption{\small An illustration of (a) listen-before-talk (LBT) and (b) the proposed consensus-before-talk (CBT) spectrum etiquettes.}
\end{figure}

\section{Related Works}

Towards enabling distributed DSA, unlicensed spectrum etiquettes have been suggested, which rely on the transmitter-side information \cite{Bianchi1996, Ziouva2002, Song2016, Ko2016, Leu2006} or on the receiver-side information \cite{Lagen18:LBR,Kim18:SAP}. These approaches are commonly rooted in random carrier-sensing techniques, and thus are still insufficient for supporting a large number of users with a strict latency guarantee due to collisions. With the aid central management, as used in Citizens Broadband Radio Service (CBRS) \cite{bi:CBRS,Souryal2014,Sohul2015} and Licensed Shared Access (LSA) \cite{bi:LSA,Buckwitz2014}, one may partly control the spectrum access requests, thereby ameliorating the collision problem. Nevertheless, due to the central controller's complexity, it is difficult to cope with dynamic user traffic, motivating this research.

Towards implementing DLT, distributed consensus algorithms have been studied under both large-scale \cite{bi:Bitcoin,Kim18:BlockFL} and small-scale systems \cite{Lamport1982, bi:pbft, Sousa2017}, with the scalability in terms of the number of participating users. In general, large-scale consensus algorithms suffer from too long consensus latency, e.g., avg. 10 min. in Blockchain \cite{bi:Bitcoin}. On the contrary, small-scale consensus algorithms, such as Byzantine fault tolerant tolerance (BFT) schemes, guarantee a fast consensus latency. Nonetheless, they cannot further increase the participating users, e.g., up to a few dozen participants \cite{Lamport1982, bi:pbft, Sousa2017}, because of the transaction exchange delays before starting the consensus process. 

To resolve this latency-scalability trade-off, directed acyclic graph (DAG) based solutions have been propose \cite{Popov:Tangle,LeMahieuLNano, bi:Hashgraph}. As opposed to the conventional Blockchain structure, each transaction in a DAG based algorithm is connected to more than one transactions so that their connections become tangled. Since a DAG has a topological ordering upon the vertices of the graph, the transaction order can readily be retrieved, thereby preventing double-spending problems. Furthermore, from the perspective of the security guarantee, Blockchain has a single-chained structure, and tries to guarantee its security level by inserting dummy computation, i.e. Proof-of-Work (PoW), in-between the vertices, i.e., blocks. On the contrary, a DAG based solution has the vertices multi-dimensionally connected to the other vertices, and this tangled structure automatically guarantees its security level, which can be beneficial for saving power.

Recently, one of the DAG based solutions, the Hashgraph algorithm \cite{bi:Hashgraph} has been proposed, which minimizes the BFT algorithm's transaction exchange bottleneck via a local consensus protocol, thereby allowing the algorithm to support far more participating users. To be specific, the Hashgraph algorithm employs a random gossip algorithm as a means of disseminating transactions. Generally, a gossip algorithm randomly selects the source-destination pairs, and spreads information in an asynchronous fashion. When it comes to its BFT application, on the one hand, it is efficient for quickly disseminating information without suffering from severe interference. On the other hand, its asynchronous dissemination may incur stragglers, and obstructs completing the dissemination, prerequisite to BFT consensus, thus bringing about too long BFT consensus latency. On this account, the Hashgraph algorithm exploits an asynchronous BFT consensus protocol that operates simultaneously with the asynchronous gossip dissemination. Inspired by this protocol, in this paper we design a local consensus algorithm for CBT.

\section{Consensus-Before-Talk (CBT) Architecture}
In this section, we describe the propose CBT architecture. For the sake of convenience, following the convention in DSA, hereafter we consider primary and secondary users, and focus primarily on the secondary users' spectrum access. With non-zero primary users, it may correspond to a WiFi-cellular coexistence scenario; otherwise, it can be interpreted as a stand-alone WiFi scenario.

As shown in Fig.~2, the CBT architecture comprises: a spectrum access transaction (SAT), a distributed spectrum ledger (DSL), and a consensus policy module. In CBT, a secondary user's access request is encapsulated in SAT and exchanged with all the other secondary users. For each received SAT, the secondary user initiates a consensus protocol with a pre-defined consensus policy. Once it reaches the consensus with all the other secondary users, the SAT is verified, and is stored in the secondary user's local DSL. Each component of the said consensus process is detailed in the following subsections.

\subsection{Spectrum Access Transaction (SAT)}
Each secondary user generates a single SAT when the user requests spectrum access. At the SAT, the user records its generation timestamp and the corresponding digital signature created by a public-key algorithm, e.g., Rivest-Shamir-Adleman (RSA) cryptosystem \cite{RSA:03}. Then, the SAT is exchanged with all the other secondary users by using a gossip protocol. For each received SAT, the secondary user first verifies the digital signature by the public key of the SAT generator, and then adds the record of its verified timestamp. As a result, each SAT contains (i) a single generated timestamp and (ii) verified timestamps cumulatively recorded by the secondary users. Note that any verified timestamp is larger than the generated timestamp, due to the propagation~delays.

\subsection{Distributed Spectrum Ledger (DSL)}
Each secondary user possesses a DSL. As shown in Fig.~2, the DSL consists of spectrum access queue (SAQ), spectrum access history (SAH), a consensus policy module, and a header. The SAQ is a queue of the SATs, and is managed by the consensus policy module that enables the consensus process and adjusts the scheduling priority of the consensus SATs. The arrival rate of the SAQ is determined by the secondary user traffic  as well as by the gossip and consensus delays in CBT. The service rate of the SAQ is set as the number of maximum accessible secondary users for a unit timespan that is hereafter given as a number $\mu$ of time slots. The maximum accessible amount is determined by the primary user access information stored in the header. This information is periodically updated by an external spectrum sensing entity, as done in Citizens Broadband Radio Service (CBRS) \cite{bi:CBRS,Souryal2014,Sohul2015} and Licensed Shared Access (LSA) \cite{bi:LSA,Buckwitz2014}. The served SATs are kept stored on the SAH that can affect the consensus policy as detailed next.

\begin{figure}
\includegraphics[width=\columnwidth]{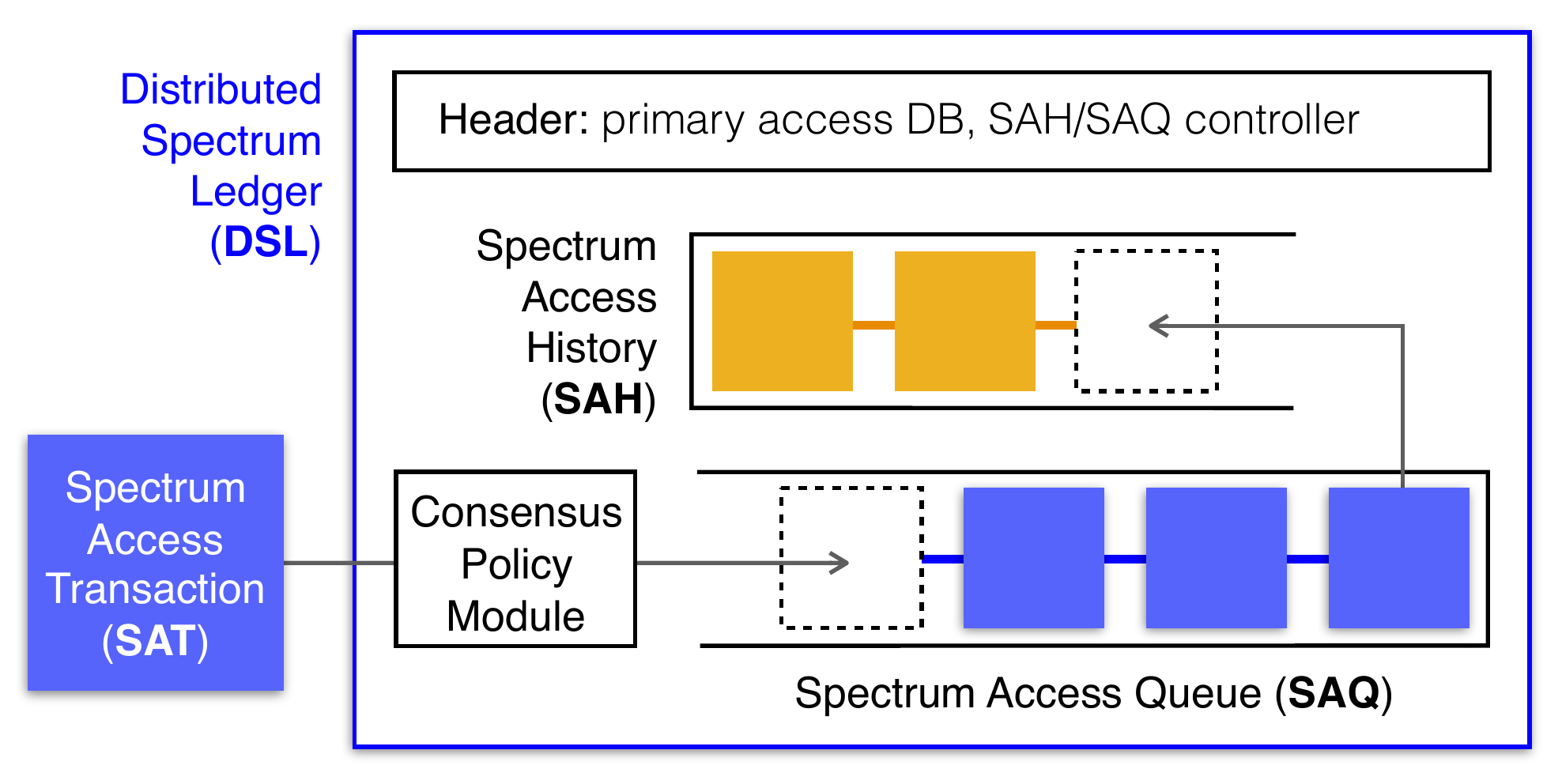}
\caption{\small An illustration of a Distributed Spectrum Ledger (DSL) that contains spectrum access queue (SAQ) and spectrum access history~(SAH).}
\label{fig:DSL}
\end{figure}

\subsection{Consensus Policy}

The consensus algorithm of CBT is based on practical Byzantine fault tolerance \cite{bi:pbft}, thus ensuring the following conditions:
\begin{itemize}
\item Termination (Liveliness) -- All SATs will be eventually known by all the secondary users;
\item Validity (Correctness) -- Invalid SATs cannot be validated by the secondary users; and
\item Agreement (Consistency) -- Two secondary users should not have disagreement on the validity and the time order of SATs.
\end{itemize}
For simplicity, we henceforth assume that all the secondary users are honest, while neglecting Byzantine users that obstruct the consensus process. The impact of the Byzantine users on CBT is to be elaborated in Section~\ref{sec:discussion}.

In CBT, the consensus objective is to enable secondary users to follow a pre-defined scheduling rule in a distributed way. To this end, at first each secondary user's spectrum access request becomes associated with its verified timestamps recorded in its corresponding SAT. Such SATs are then exchanged and verified by all secondary users. After the verification, the accumulated verified timestamps reach a consensus, following a pre-defined consensus algorithm stored in the consensus policy module within each DSL. 

Motivated by the gossip-of-gossip protocol in Hashgraph \cite{bi:Hashgraph}, the CBT consensus algorithm is locally operated at each secondary user. To elaborate, as exemplified in Fig.~3, consider user~1 generated an SAT, and the SAT is propagated through the following order: users~2$\rightarrow$3$\rightarrow$1$\rightarrow$2$\rightarrow$3. Denoting as $t_{i}(i)$ and $t_{i}(j)$ with $j\neq i$ the generated timestamp of user~$i$ and its verified timestamp by user~$j$, respectively, the generated SAT's consensus timestamp~$\hat{t}_i(j)$ at user~$j$ is given as $\hat{t}_i(j)=\sum_{k\neq j} t_{i}(k) /n$, where $n$ is the number of secondary users. Here, the consensus timestamp calculation at each user neglects its own verified timestamp, in order to avoid any selfish manipulation. So long as $n$ is sufficiently large, the consensus timestamps for different users become almost identical. Thus, each user can independently and locally calculate its own consensus time, while achieving global consensus.

The SAT containing the consensual timestamp is then stored in the SAQ of each user, according to a pre-defined scheduling rule. Two possible scheduling examples are described as below.

\begin{enumerate}
	\item \emph{First verified, first served} -- The spectrum access priority is determined by the verification timestamp order. For instance, if $\hat{t}_1(j)\leq \hat{t}_2(j)$, then $\mathrm{SAT}_1$ is placed prior to $\mathrm{SAT}_2$ in the SAQ of user~$j$.

	\item \emph{Fairness guarantee} -- Exploiting the SAH, one can maximize the fairness by prioritizing the access requests from the least served users. In a similar way, one may avoid selfish users occupying excessive spectrum bands by first counting their number of access requests in the SAH and then adjusting their next SAQ priority.

\end{enumerate}

\begin{figure}
\includegraphics[width=\columnwidth]{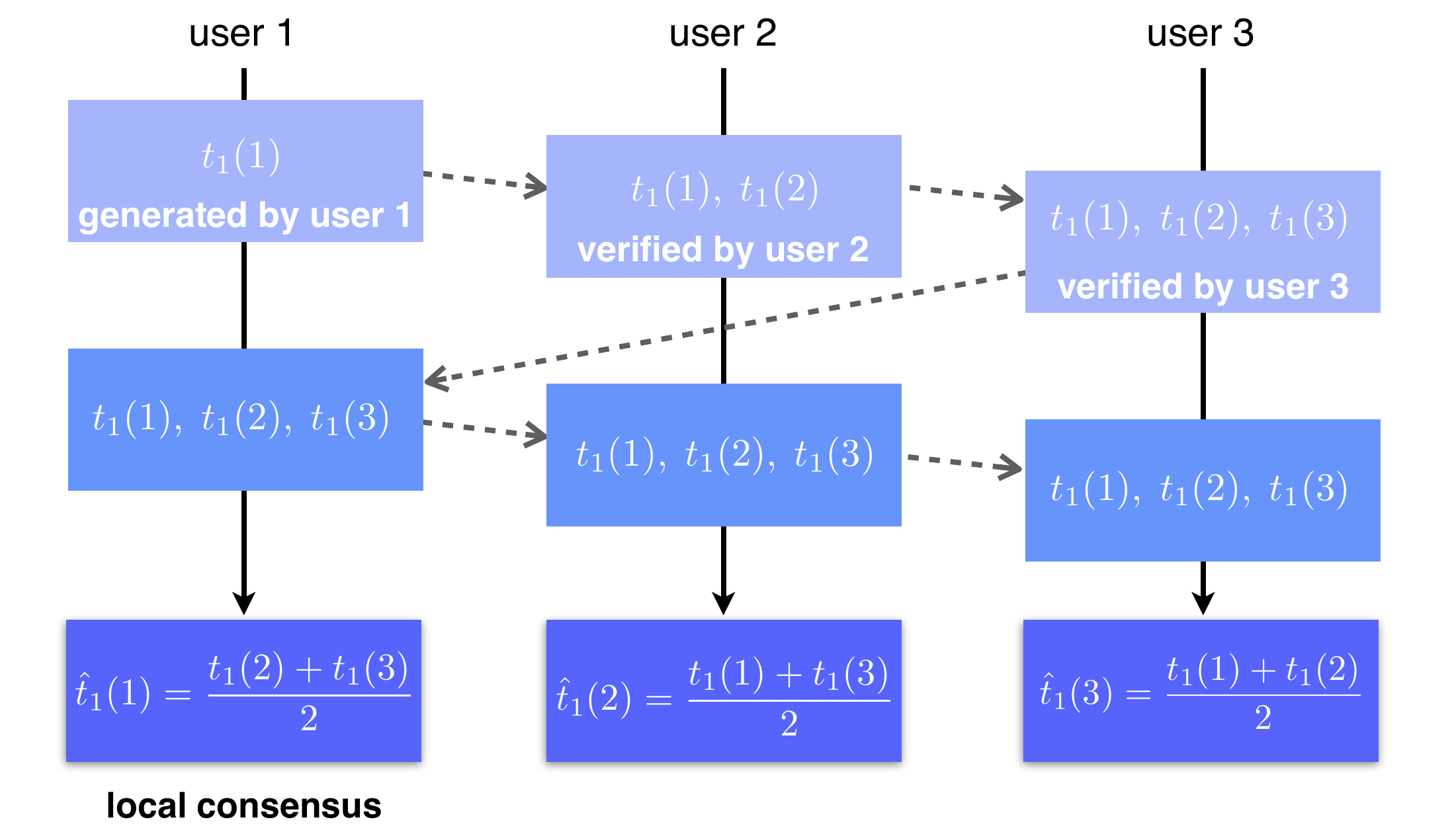}
\caption{\small An example of the consensus algorithm in CBT. For the SAT generated by user~$1$, it is propagated to users~2$\rightarrow$3$\rightarrow$1$\rightarrow$2$\rightarrow$3. Afterwards, each user~$j$ independently computes its local consensus timestamp~$\hat{t}_1(j)$.}
\end{figure}

\section{End-to-End Latency Analysis}
Following the CBT architecture proposed in Sect.~II, in this section we derive an analytic expression of the end-to-end latency under CBT, as well as under its benchmark LBT. The end-to-end latency is determined as the average delay from a secondary user's access request generation to its successful access, to be formally defined with a proper network model in the following subsections.

\subsection{Network model}
The network under study consists of a number $n$ of secondary users that are sharing the frequency-time resource with primary users. During a unit time span, set as $\mu$ time slots, a number $n_r\leq n$ of the secondary users request their spectrum access. Assuming the primary user traffic information is known from a database server as done in \cite{bi:CBRS,Souryal2014,Sohul2015,bi:LSA,Buckwitz2014}, these secondary users can access up to a number $n_v$ of vacant resource blocks during the unit time span. 

For the sake of convenience, we consider each user access consumes a single resource block, and focus only the case $n_r\leq n_v$ with a uniformly randomly selected secondary user, referred to as a typical user. The typical user generates its access request at time $t_0\in(i \mu, (i+1)\mu]$ with $i>0$. For CBT, the typical user's end-to-end latency $T_\text{CBT}$ incorporates the consensus delay. For LBT, on the contrary, the end-to-end latency $T_\text{LBT}$ includes all the back-off delays until the first successful access.

\subsection{LBT Latency}
The end-to-end latency under the benchmark LBT scheme is evaluated as follows. We consider a large number of secondary users having limited listening coverages, leading to collisions due to their hidden node problem \cite{Bianchi1996, Ziouva2002} and simultaneous spectrum access attempts. Each access collision consumes a constant back-off delay set identically as the unit time span, i.e., $\mu$ time slots. Assuming that each secondary user uniformly randomly selects its access resource block out of the total $n_v$ blocks, the typical user avoids any collision with probability $(1-1/n_v)^{n_r - 1}$.

Due to the backed-off secondary uses, at $t=i\mu$, an average number $\bar{n}_{r,i}$ of secondary users attempt spectrum access. Its sequence $[\bar{n}_{r,i}]$ is an increasing sequence that satisfies the following relation.
\begin{align}
\bar{n}_{r,i+1} = n_{r} + \bar{n}_{r,i} \left(1 -  \left(1 - 1/n_v\right)^{\bar{n}_{r,i} - 1} \right),
\end{align}
where the initial value equals $\bar{n}_{r,1} = n_r$. 

According to the fixed-point theorem \cite{Zorich}, if $n_r$ satisfies the condition 
\begin{align}\label{eq:converge}
n_r \leq -\frac{1}{e (1-1/n_v)\log(1-1/n_v)},
\end{align}
then as $i \rightarrow \infty$, $[\bar{n}_{r,i}]$ converges to a certain fixed point $\hat{n}_r$ that is the smallest root of the equation $x(1-1/n_v)^{x-1} - n_r = 0$. In this case, for some $\bar{n}_{r,i+1}$, the access success probability becomes $(1-1/n_v)^{\bar{n}_{r,i+1} - 1}$. By supposing that the number of access failure follows a geometric distribution with the parameter given by the access success probability, the typical user's average aggregate back-off delay until the first access success is given as
\begin{align} \label{Eq:LBT}
T_\mathrm{LBT} = \left( \frac{1}{\left(1- 1/ n_v\right)^{\hat{n}_r-1}} - 1 \right)\mu  + \frac{\mu}{2} ,
\end{align}
where ${\mu}/{2}$ comes from the average waiting time for the first access attempt after the generation of an access request.

On the other hand, if $n_r$ does not satisfy the condition \eqref{eq:converge}, then $[\bar{n}_{r,i}]$ diverges as $i \rightarrow \infty$. It implies that the number of collisions keep increasing over time, and eventually any secondary user access becomes unavailable. Finally, combining this with \eqref{eq:converge} and \eqref{Eq:LBT}, we obtain the typical user's end-to-end latency under LBT as:
\begin{align}\label{eq:T_LBT}
T_{\mathrm{LBT}} = 
  \begin{cases} 
\frac{\mu}{\left(1- 1/ n_v\right)^{\hat{n}_r-1}}  - \frac{1}{2}\mu & \mathrm{if\ (\ref{eq:converge})}\\ 
\infty       & \text{otherwise}. 
  \end{cases}
\end{align}

\subsection{CBT Latency}\label{subsec:cbtlatency}

Assuming the delays incurred by the local consensus time stamp calculations addressed in Sect.~II-C are negligibly small, the consensus latency for the typical user is given by the delay brought by the typical user's disseminating its SAT to all the other secondary users. For the SAT dissemination, we consider a \emph{push gossip} algorithm \cite{Boyd2005}, where the SAT is transmitted to a randomly chosen target user, regardless of whether the target user has already received the SAT or not. During the SAT dissemination process, we neglect their interference and collisions in that the gossip algorithm can easily mitigate the concurrent transmissions within a small region.

With the push gossip algorithm at time $t \geq t_0$, our focus is to derive the dissemination delay $t-t_0$ such that the average number $n_s(t)$ of users who received the typical user's SAT becomes the entire $n-1$ users. To this end, following \cite{Bailey1975, Demers1987}, $n_s(t)$ is given by a logistic difference equation:
\begin{align}
n_s(t+1) = n_s(t) + \phi n_s(t)\left(1-\frac{n_s(t)}{n} \right). \label{Eq:logistic}
\end{align}
where $\phi$ denotes the number of receivers that can be concurrently connected to a single transmitter; e.g., $\phi =1$ implies a one-to-one pairwise communication, while $\phi>1$ indicates one-to-many broadcast communication. Applying the inverse of the Euler's approximation, \eqref{Eq:logistic} in discrete time domain is recast as the following differential equation in continuous time domain:
\begin{align}\label{eq:gossipdiff}
\frac{d}{dt}n_s(t) = \phi n_s(t) \left(1-\frac{n_s(t)}{n}\right),
\end{align}
with the initial condition $n_s(t_0) = 1$. By solving \eqref{eq:gossipdiff}, we obtain
\begin{align}\label{eq:infected}
n_s(t) = \frac{n}{1+(n-1)e^{- \phi (t - t_0)}}.
\end{align}

In \eqref{eq:infected}, it reads $\lim_{t\rightarrow \infty} n_s(t)= n$, thus asymptotically guaranteeing the termination condition in Sect.~II-C. In order to derive non-asymptotic delay, we suppose the dissemination of an SAT becomes completed at $t$ if a fraction $\gamma=n_s(t)/n<1$ of the users receive the SAT, where the target gossip success proportion $\gamma$ is set as the value close to $1$, e.g., 0.999. Then, the corresponding dissemination delay $t-t_0$ of the typical user's SAT equals
\begin{align}
t-t_0 =  \frac{1}{\phi}\log\left(\frac{1+(n-1)\gamma}{1-\gamma} \right). \label{Eq:SingleDiss}
\end{align}

Note that \eqref{Eq:SingleDiss} is the SAT dissemination delay. As addressed in Sect.~II-C, this does not guarantee to achieve the consensus, which requires one additional round of the dissemination. Incorporating this, we finally obtain the end-to-end latency under CBT:
\begin{align}
T_{\mathrm{CBT}} =  \frac{2 n_r}{\phi}\log\left(\frac{1+(n-1)\gamma}{1-\gamma} \right) + \frac{\mu}{2},\label{Eq:Tcbt}
\end{align}
where the first multiplication term $n_r$ in \eqref{Eq:Tcbt} is because there exist $n_r$ spectrum access requesting secondary users for every $\mu$ time slots. The term ${\mu}/{2}$ comes from the average waiting time of the SAQ.

\section{Simulation results}
In this section we numerically evaluate the effectiveness of the proposed CBT. A random push gossip protocol is used for CBT in all simulations and we assume that all communications during gossip algorithm is pairwise, i.e., $\phi = 1$. All simulation results are the averaged output of 10,000 iterations.

Fig. \ref{fig:gamma} describes the time required to disseminate a SAT to the secondary users, i.e., gossip delay, versus the fraction of secondary users who received SAT. The number of secondary users is fixed $n = 1,000$ and only a single SAT is generated in the network, If the fraction $\gamma = 0.99$, it means that $990$ secondary users received the SAT. First of all, the figure shows that the gossip algorithm analyzed in \ref{subsec:cbtlatency} is precise in some degree, when $\gamma <0.996$. On the other hand, if $\gamma \geq 0.996$, the gap between the simulation and analysis increase and the analysis of gossip delay tends to diverge eventually. From the simulation, the gossip delay of a complete gossiping, i.e., $\gamma = 1$ is around $14.5$. The delay is almost same as the gossip delay obtained by analysis when $\gamma = 0.999$. Therefore, we fix $\gamma = 0.999$, i.e., the 99.9\% of the secondary users successfully exchange all the SATs during $\mu$ time slots, in the rest of the experiments.

Fig. \ref{fig:LBTCBT} shows the latency performance comparison between LBT and the proposed CBT protocol, with respect to the number of secondary access requests in $\mu$ time slots, that is $n_{r}$. The latency is normalized by $\mu = 1,000$, $5,000$ and $10,000$, and the number of secondary users and the number of vacant resource blocks in every $\mu$ time slots is fixed to $n = 1,000$ and $n_v = 100$, respectively. Note that $\mu$ is closely related to the ratio between the size of SAT and the information size communicated via a single access by a primary or secondary user. Clearly, the normalized latency of LBT is independent of $\mu$, however, the normalized latency becomes smaller as $\mu$ increases in the proposed CBT. For $\mu = 5,000$ and $10,000$, the normalized latency of CBT is small compared to that of LBT. On the other hand, when $\mu = 1,000$, LBT outperforms CBT, which means that the consensus process in CBT requires more than $1,000$ time slots. However, the latency of LBT increases exponentially as $n_r$ increases and near $n_r = 34$, there is a crossing point and CBT outperforms LBT. For $n_r \geq 36$, the normalized latency of LBT diverges, since from ({\ref{eq:T_LBT}), LBT cannot serve more than $36$ secondary accesses in $\mu$ time slots due to collisions between the secondary access requests.

Fig. \ref{fig:scalability} shows the latency performance comparison between LBT and CBT, with respect to the number of secondary users in the network. In order to show the case when there exists a crossing point on the latency performance of LBT and CBT as the number of secondary users increases, we fix the numbers $n_r = 10$ and $\mu = 2,500$ in this experiment. In the figure, it is shown that the normalized latency of CBT increases logarithmically with respect to the increase of the number of secondary users, while the latency of LBT remains unchanged and unaffected by the increase of the secondary users. This is because the required time for a consensus process in CBT is largely dependent on the number of participating users in the consensus process, while LBT does not have this kind of process. This scalability issue will be discussed in the later section and in the future works.

\begin{figure}
\centering
\includegraphics[width=\columnwidth]{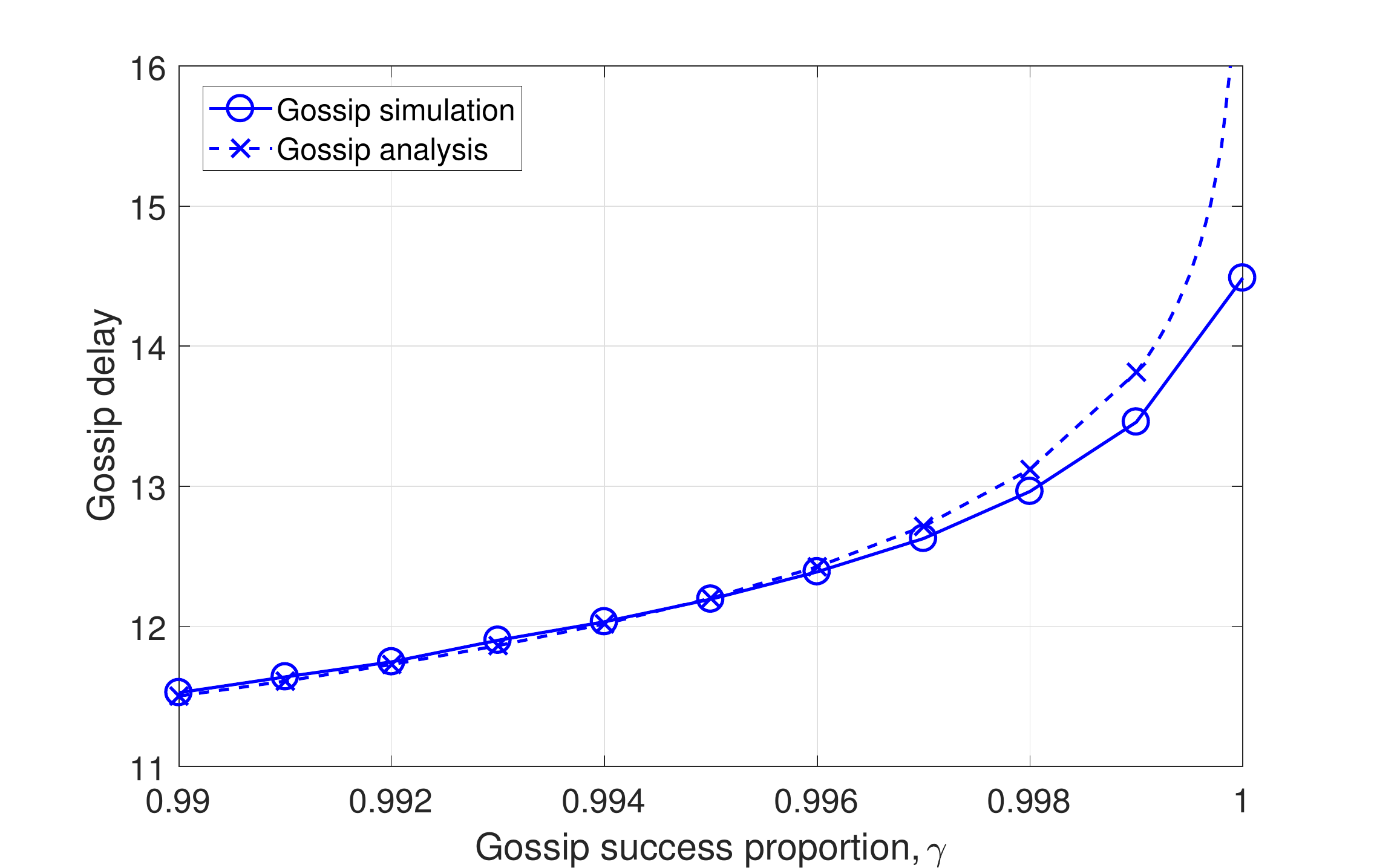}
\caption{\small Gossip delay versus the gossip success proportion.}
\label{fig:gamma}
\end{figure}

\begin{figure}
\centering
\includegraphics[width=\columnwidth]{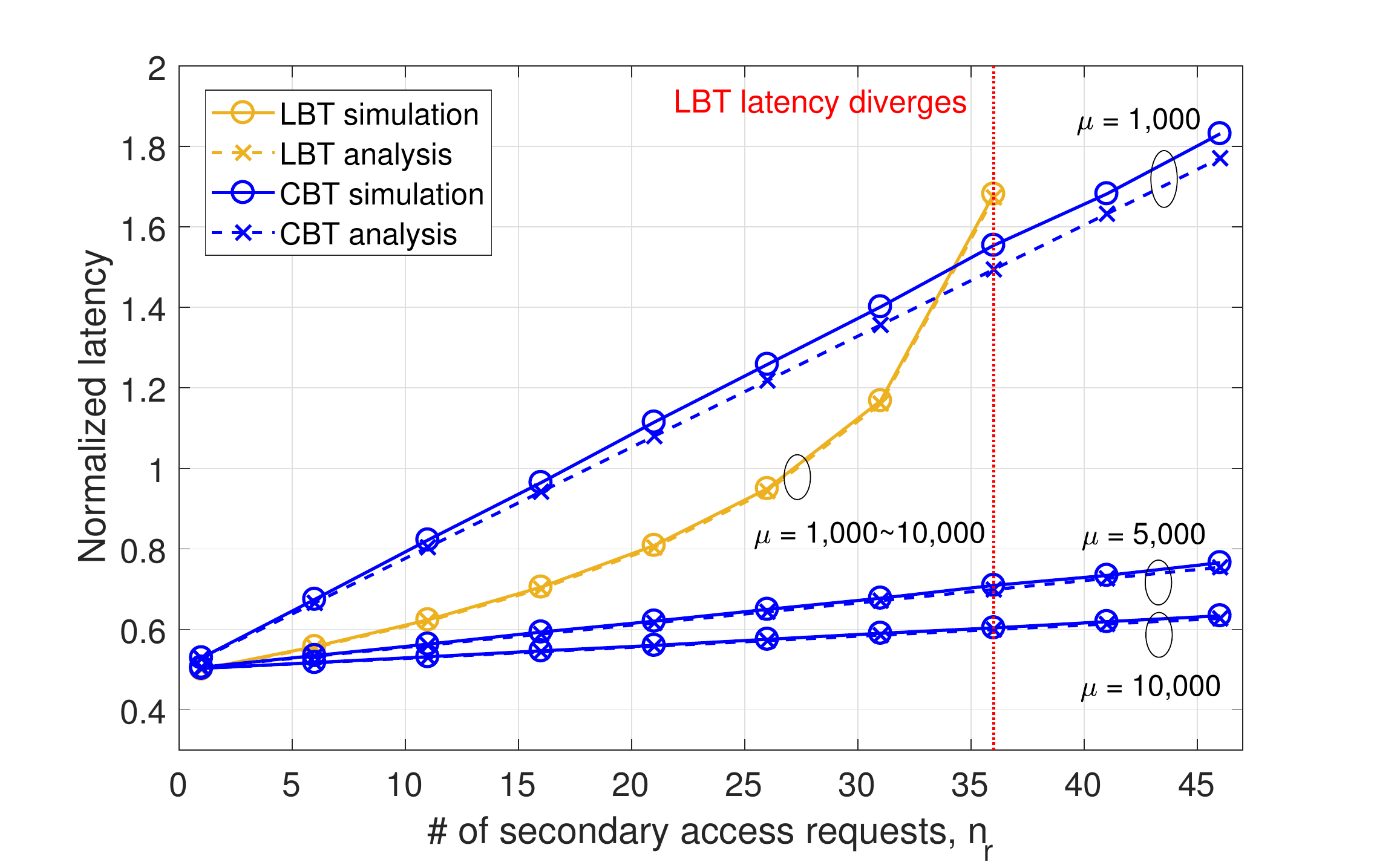}
\caption{\small Latency normalized with $\mu = 1,000$, $5,000$ and $10,000$ versus the number of secondary access requests in $\mu$ time slots for LBT and the proposed CBT.}
\label{fig:LBTCBT}
\end{figure}

\begin{figure}
\centering
\includegraphics[width=\columnwidth]{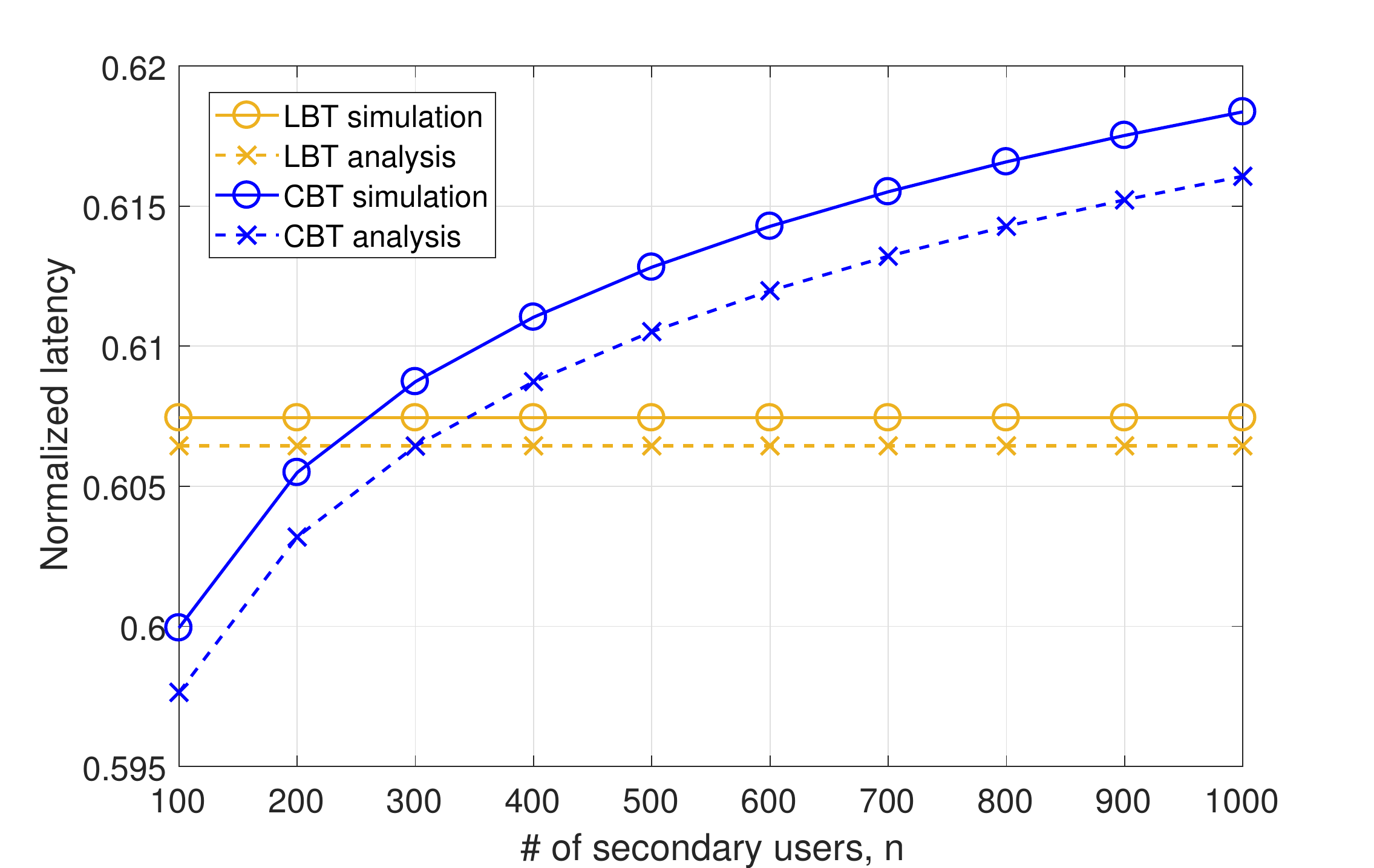}
\caption{\small Latency normalized with $\mu = 2,500$ versus the number of secondary users for LBT and CBT.}
\label{fig:scalability}
\end{figure}

\section{Discussion}\label{sec:discussion}
In order to further improve the proposed CBT spectrum etiquette architecture, this section describes 
(i) how many users are need to participate in, (ii) how to exchange their SATs, and (iii) how to calculate the consensus timestamps, as follows.

\subsection{Consensus Participation -- Direct vs. Representative}
Reaching a consensus with many users are more tolerant to the various attacks compared to reaching a consensus with small number of users. However, if there is too many nodes in the consensus process, it will cause a critical latency problem, since the time required to reach a consensus increases as the number of nodes increases. Hopefully, if there are a set of nodes, so called \emph{representative nodes}, that can operate a consensus process as a representative, the secondary users can have a common result on the spectrum scheduling in a short period time. In other words, depending on how many of the representative nodes are chosen, there will be a trade-off between the consensus latency and the security performance. Therefore, for a given required condition for latency and security, we will investigate and discuss about the optimal number of representative consensus nodes in our future works.

\subsection{Gossip Protocol -- Push vs. Pull}
The gossip algorithm considered in this paper is push gossip. In push gossip, the users who have the gossip message randomly selects the receivers and pushes the message to the target. As the time goes by, the number of users who does not have the message will become smaller, however, disseminating speed of the gossip message will slow down since the receivers are randomly selected among all the users and the portion of the users who have not yet got the message becomes smaller. On the other hand, in \emph{pull gossip}, the users who have the message makes use of a side information obtained from the users who have not yet got the message. In other words, the users who have not yet got the message requests the gossip message from the message holders. This makes the message holders to disseminate the message by choosing the random users among the smaller set of users. However, in order to deliver the side information, it costs extra communication resources.  Therefore in order to implement an efficient gossip algorithm in our structure, mixing two types of gossip algorithm can be one solution. For example, at the beginning, the gossip message holders use push gossip to disseminate the message. After some time, the disseminating speed will be slowed down and  then the gossip algorithm can be changed to pull gossip to speed up the dissemination.

\subsection{Consensus Timestamp Calculation -- Mean vs. Median}
In the previous section, we assumed that all the secondary users are honest. However, in the real world environment, a selfish secondary can be even more selfish so that they cheat during the consensus process. For example, Alice and Bob generate $\mathrm{SAT}_a$ and $\mathrm{SAT}_b$, respectively, at time $t = 0$ and $\mathrm{SAT}_a$ is verified by Bob, Carol and David at $t = 2$. Meanwhile, $\mathrm{SAT}_b$ is verified by Carol and David at $t =1$ and if Alice is a cheater and selfish so that she wants to put her transaction into the queue before that of Bob, she can simply delay verifying $\mathrm{SAT}_b$ and stamp time at, say $t = 98$. Then the computed average timestamp of $\mathrm{SAT}_a$ and $\mathrm{SAT}_b$ will be $2$ and $50$, respectively, which puts $\mathrm{SAT}_a$ in a higher priority.

Meanwhile, rather than taking the average timestamp as the representative value for the distributed consensus, using the median timestamp in the consensus process is more tolerant to Byzantine attacks. In the above example, if the median timestamp is taken to decide access priority of the users, then Bob accesses before Alice since the median timestamp of $\mathrm{SAT}_a$ and $\mathrm{SAT}_b$ is $2$ and $1$, respectively, which is tolerant to the selfish attack by Alice.

\section{Conclusion}
This paper proposes a CBT spectrum etiquette based on the distributed spectrum ledger technology. We introduced the structure and mechanism of CBT and also analyzed it from a technical point of view. Specifically, a latency performance is compared with the conventional LBT and showed that under high secondary traffic environment, the proposed CBT performs better since it avoids collisions via distributed consensus based spectrum scheduling.

\section*{Acknowledgment}
This work was supported in part by the Academy of Finland project CARMA, and 6Genesis Flagship (grant no. 318927), in part by the INFOTECH project NOOR, in part by the Kvantum Institute strategic project SAFARI, and in part by the Korea Electric Power Corporation (grant no. R17XA05-63).



\bibliographystyle{ieeetr}  
\bibliography{IEEEabrv,Blockchain}

\end{document}